\documentclass[aps,prb,amsmath,amssymb,twocolumn,superscriptaddress,floatfix]{revtex4-1}
\usepackage{helvet}
\usepackage{graphicx}
\usepackage{hyperref}
\hypersetup{pdftitle={Al6Re},pdfauthor={Darren C. Peets},pdfsubject={Al6Re},pdfdisplaydoctitle}

\def\fourRe{Al\ensuremath{_4}Re}
\def\sixRe{Al\ensuremath{_6}Re}
\def\Tc{\ensuremath{T_\text{c}}}
\def\Hc{\ensuremath{H_\text{c}}}
\def\Hc2{\ensuremath{H_\text{c2}}}
\begin{document}

\title{Type-{I} superconductivity in {Al$_6$Re}}

\author{Darren C. Peets}
\email{dpeets@nimte.ac.cn}
\affiliation{State Key Laboratory of Surface Physics, Department of Physics; and Advanced Materials Laboratory, Fudan University, Shanghai 200438, China}
\affiliation{Ningbo Institute of Materials Technology and Engineering, Chinese Academy of Sciences, Ningbo, Zhejiang 315201, China}

\author{Erjian Cheng}
\affiliation{State Key Laboratory of Surface Physics, Department of Physics; and Advanced Materials Laboratory, Fudan University, Shanghai 200438, China}
\author{Tianping Ying}
\affiliation{State Key Laboratory of Surface Physics, Department of Physics; and Advanced Materials Laboratory, Fudan University, Shanghai 200438, China}
\altaffiliation[Current address:]{Laboratory for Materials and Structures, Tokyo Institute of Technology, Yokohama 226-8503, Japan}

\author{Markus Kriener}
\affiliation{RIKEN Center for Emergent Matter Science (CEMS), Wako 351-0198, Japan}

\author{Xiaoping Shen}
\affiliation{State Key Laboratory of Surface Physics, Department of Physics; and Advanced Materials Laboratory, Fudan University, Shanghai 200438, China}

\author{Shiyan Li}
\email{shiyan\_li@fudan.edu.cn}
\affiliation{State Key Laboratory of Surface Physics, Department of Physics; and Advanced Materials Laboratory, Fudan University, Shanghai 200438, China}
\affiliation{Collaborative Innovation Center of Advanced Microstructures, Nanjing 210093, China}
\author{Donglai Feng}
\email{dlfeng@fudan.edu.cn}
\affiliation{State Key Laboratory of Surface Physics, Department of Physics; and Advanced Materials Laboratory, Fudan University, Shanghai 200438, China}
\affiliation{Collaborative Innovation Center of Advanced Microstructures, Nanjing 210093, China}
\affiliation{Hefei National Laboratory for Physical Science at the Microscale and Department of Physics, University of Science and Technology of China, Hefei, Anhui 230026, China}

\begin{abstract}

While the pure elements tend to exhibit Type-I rather than Type-II superconductivity, nearly all compound superconductors are Type-II, with only a few known exceptions.  We report single crystal growth and physical characterization of the rhenium aluminide \sixRe, which we conclude is a Type-I superconductor based on magnetization, ac-susceptibility, and specific-heat measurements.  This detection of superconductivity, despite the strong similarity of \sixRe\ to a family of W and Mo aluminides that do not superconduct, suggests that these aluminides are an ideal testbed for identifying the relative importance of valence electron count and inversion symmetry in determining whether a material will superconduct.  

\end{abstract}

\maketitle

\section{Introduction}

A vast array of exotic physics is possible in noncentrosymmetric superconductors, which prompted our prior project probing a family of tungsten and molybdenum aluminides\,\cite{PeetsPRM2018}.  Nearly all members of this family are noncentrosymmetric, most are straightforward to grow from aluminum flux, and all are excellent metals; however, none were found to superconduct down to 100-300\,mK.  One possible culprit here is electron count.  One of Bernd Matthias' better-known rules for finding new superconductors states that having six valence electrons per transition metal atom (as in metallic tungsten and molybdenum) is the least optimal, while five or seven electrons provide the highest transition temperatures\,\cite{Matthias1955,Collver1973,Geballe2015,MatthiasRules}.  If we thus gloss over the fact that these aluminides are almost entirely aluminum, substituting a transition metal one column to the left or right in the periodic table ought to greatly improve our chances of finding superconductivity.  Similar cage aluminides are not observed for V, Nb, or Ta one column to the left, but they are for Mn, Tc, and Re to the right.

Since finding exotic physics in noncentrosymmetric superconductors requires a significant spin-splitting of the bulk band structure, which in turn requires the strong spin-orbit coupling associated with high atomic number, we turned to the Al--Re system.  As with the Al--W and Al--Mo phase diagrams, the Al--Re phase diagram includes a cascade of Al-rich phases\,\cite{Huang1998,Cornish1999,Schuster2001,Balanetskyy2008}, although in the Re case many of them are centrosymmetric.  No low-temperature physical properties of any of these phases have been reported.  We targeted the high- and low-temperature phases of \fourRe, for the most direct comparison to our previous work.  However, the crystals we obtained were centrosymmetric \sixRe.  Based on magnetization, ac susceptibility, and specific heat measurements, we conclude that this material is a Type-I superconductor with a transition temperature of 0.74\,K and a critical field of roughly 50\,Oe.  \sixRe\ joins a very small number of compounds known to be Type-I superconductors\,\cite{Soulen1977,VanVechten1983,Kobayashi1981,Palstra1986,Yamaguchi1987,Zhao2012,Gottlieb1992,Yonezawa2005,Ren2007,Muranaka2008,Kriener2008,Kriener2008b,Svanidze2012,WangSnAs2014,Sun2016,Rebar2019,Singh2019,Beare2019,Hull1981,Anand2011}.
\section{Experimental}

\begin{figure*}
  \includegraphics[width=\textwidth]{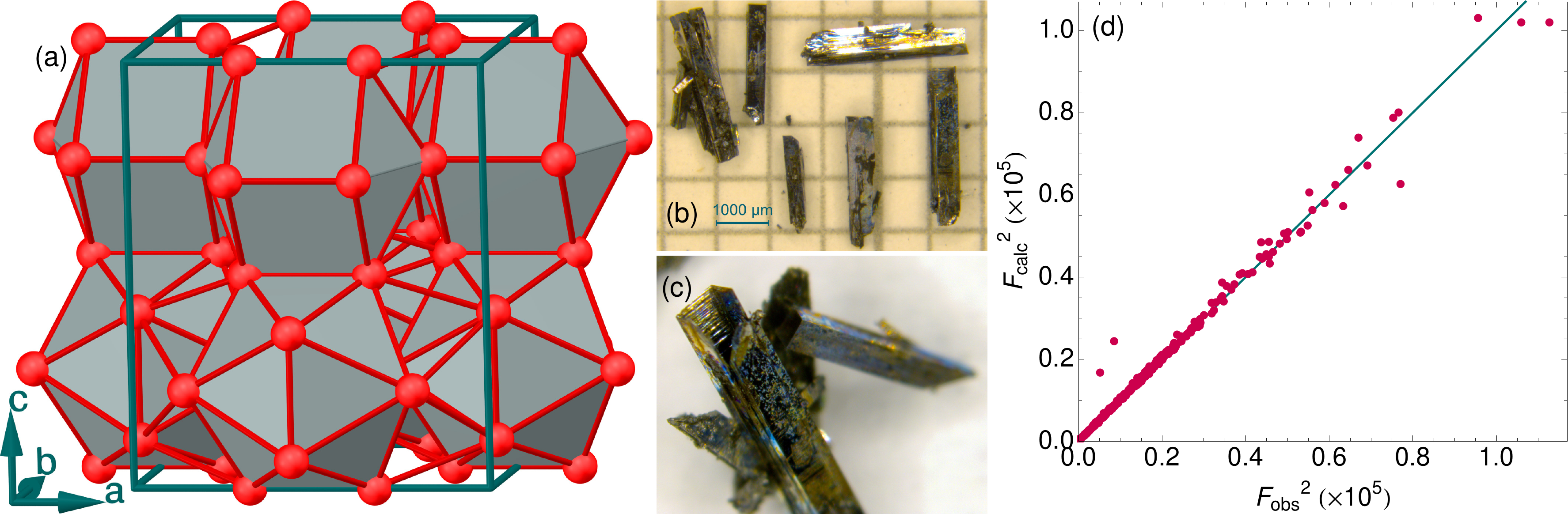}
  \caption{\label{fig:XRD}Structure of \sixRe.  (a)~Refined crystal structure --- the Re atoms are located at the centres of the Al cages. (b,c)~Photos of several representative \sixRe\ crystals (b)~on mm-ruled graph paper, and (c)~demonstrating the morphology. (d)~Quality of the structure refinement.}
\end{figure*}

Al wire (PRMat, 99.999\%) and rhenium powder (Alfa Aesar, 99.99\%), in the approximate molar ratio 200:1, were weighed into alumina crucibles inside an Ar-filled glovebox.  These crucibles, with Al$_2$O$_3$ lids to limit damage to the quartz by Al vapour, were then sealed under vacuum inside quartz tubes.  The solubility of Re in Al is low at the relevant temperatures\,\cite{Huang1998,Cornish1999,Schuster2001,Balanetskyy2008}, so the molar ratios were chosen to limit the volume of crystals rather than to reach the liquidus line.  The original intent was to grow the low-temperature phase of \fourRe, which is accessible from the melt between peritectic melting points at roughly 813 and 1000\,$^\circ$C, so the crucible was cooled from 975 to 830\,$^\circ$C over the course of about 5 days, after first spending 3\,h at 975\,$^\circ$C to melt the contents.  At the conclusion of growth, the furnace was switched off and allowed to cool freely.  The Al flux was dissolved off in 1\,M HCl, revealing two types of crystals, likely due to a lower final temperature than intended.  Large, rectangular blocks encased in smaller crystals were not found to superconduct down to 300\,mK and were not investigated in detail.  This is most likely the low-temperature phase of \fourRe, also known as Al$_{33-x}$Re$_8$.  The second type of crystal was shiny, black needles of size up to $\sim$1$\times$1$\times$6\,mm$^3$ as shown in Fig.~\ref{fig:XRD}b.  These were actually hollow square tubes with one open end (Fig.~\ref{fig:XRD}c).


Single-crystal x-ray diffraction was performed on a 0.38$\times$0.12$\times$0.10\,mm$^3$ crystal using a Bruker D8 Venture diffractometer with an APEX-II CCD area detector and a molybdenum K$\alpha$ source.  The structure was solved and refined using the SHELX suite of software\,\cite{Sheldrick2015}.  Stoichiometries were verified by electron-probe microanalysis (EPMA), using a Shimadzu EPMA-1720, with beam current 10\,nA accelerated at 15\,kV.  Standard samples were the pure elements, using the Al K$\alpha$ and Re M$\alpha$ lines analyzed using a RAP (rubidium acid phthalate) crystal for Al, and PET (pentaerythritol) for Re.  Resistivity was measured for fields along [110] in an Oxford $^3$He cryostat or using a Quantum Design Physical Property Measurement System (PPMS) as a cryostat, with the data collected using an external lock-in detector. Currents (3\,mA for high-temperature data, 10\,$\mu$A to observe superconductivity) were along the needle axis, which is [001]. A separate current source was used and an additional resistance of several kilohms was added in series with the current leads, to enhance measurement stability.  To extract a low-temperature power law, an offset was fit and subtracted, a log-log plot was used to identify an appropriate upper temperature limit for the fit, then finally a least-squares fit to $\rho(T) = \rho_0 + AT^\alpha$ was performed to the original data below that temperature limit.  

Magnetization was measured in a Quantum Design Magnetic Properties Measurement System (MPMS3) equipped with an iHelium3 $^3$He refrigerator, with crystals affixed to a straw and the needle axis, [001], parallel to the field.  A small field-dependent magnetization offset attributed to a film of chloride hydrates on the surface was subtracted from the temperature-dependent data, and a small linear contribution was subtracted from the field-dependent data.  ac Susceptometry was measured in a PPMS with the ACDR (dilution refrigerator ac susceptibility) option, using an excitation amplitude of 0.1\,Oe at 9984\,Hz; mosaics of crystals were mounted to the side of a sapphire bar using GE varnish.  A small diamagnetic offset was subtracted by assuming zero sample contribution well above the zero-field \Tc, and a small phase offset was corrected by assuming zero loss from the sample at temperatures far below \Tc\ at low field.  Low-temperature specific heat was measured on several mosaics of crystals between 0.1 and 5\,K using a PPMS with the $^3$He or dilution refrigerator option and the relaxation time method.  Crystals were mounted using Apiezon N grease, with the field applied along [110].  Higher-resolution data near the transition required much weaker heat pulses and the averaging of more measurements to compensate for the weaker signal.  

\section{Crystal Structure}

EPMA indicated a composition of Al$_{6.18(13)}$Re, consistent with \sixRe\ within 1.4$\sigma$.  Single-crystal x-ray diffraction similarly concluded that the crystals were \sixRe.  The refined atomic positions are presented in Tab.~\ref{tab:atoms}, while details of the refinement and the refined anisotropic displacement parameters are presented in Tabs.~\ref{tab:XRD} and \ref{tab:U}, respectively, in Appendix \ref{XRDeets}.  The refined crystal structure is shown in Fig.~\ref{fig:XRD}a, and Fig.~\ref{fig:XRD}d shows the relation between the observed and calculated intensities $F^2$, indicating a good refinement.  The material crystallizes in the centrosymmetric orthorhombic space group $Cmcm$ (No.~63), with lattice parameters $a=7.599(4)$\,\AA, $b=6.602(4)$\,\AA, and $c=9.040(5)$\,\AA\ at 203(2)\,K.  As previously reported\,\cite{Wilkinson1967,Niemann1993}, Re is contained in Al cages, and as in Al$_4$W, Al$_5$W, and Al$_{49}$Mo$_{11}$\,\cite{PeetsPRM2018}, the cages are too tight for rattling modes that would enhance superconductivity --- Al--Re bond lengths are consistent with those expected based on the elements' atomic or covalent radii.


\begin{table}[htb]
  \caption{\label{tab:atoms}Refined atomic positions for \sixRe\ in $Cmcm$ (No.\ 63) with lattice parameters $a$=7.599(4)\,\AA, $b$=6.602(4)\,\AA, $c$=9.040(5)\,\AA, and $Z$=4, at 203(2)\,K.}
  \begin{tabular}{lcr@{.}lr@{.}lr@{.}lr@{.}l}\\ \hline\hline
    Site & Mult. & \multicolumn{2}{c}{$x$} & \multicolumn{2}{c}{$y$} & \multicolumn{2}{c}{$z$} & \multicolumn{2}{c}{$U_{eq}$ (\AA$^2$)}\\ \hline
    Re & $4c$ & 0&5000 & 0&0449(1) & 0&7500 & 0&005(1)\\
    Al(1) & $8e$ & 0&3250(3) & 0&0000 & 0&5000 & 0&010(1)\\
    Al(2) & $8f$ & 0&5000 & 0&3686(3) & 0&6002(2) & 0&011(1)\\
    Al(3) & $8g$ & 0&1806(2) & 0&2096(3) & 0&7500 & 0&010(1)\\ \hline\hline
  \end{tabular}
\end{table}


\section{Physical Properties}

\begin{figure}[htb]
  \includegraphics[width=\columnwidth]{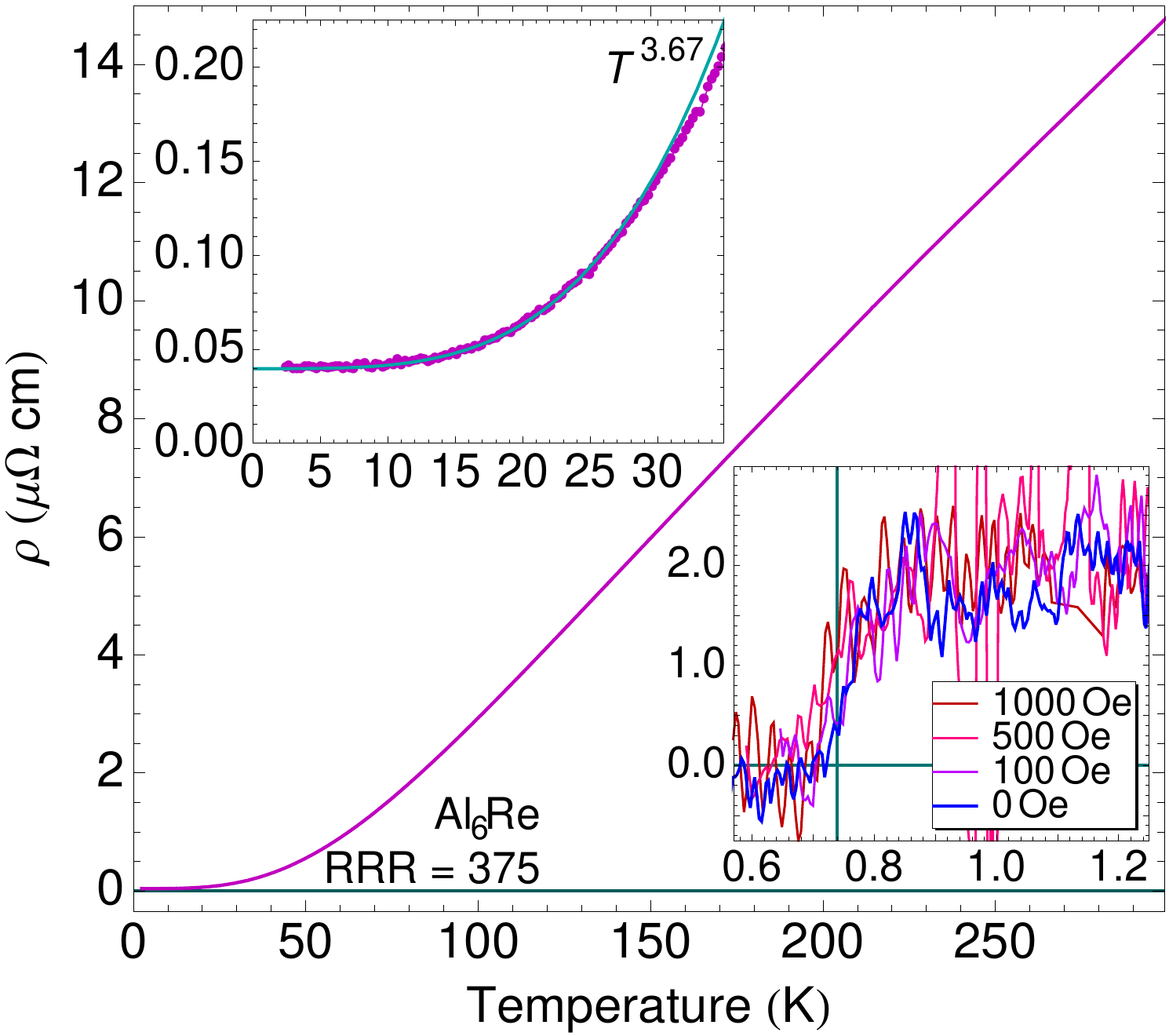}
  \caption{\label{fig:rho}Resistivity of \sixRe, with its residual resistivity ratio and low-temperature power law shown, for current along the needle axis.  The upper inset demonstrates the low-temperature power law fit.  The lower inset shows the possible superconducting transition in several fields $H\parallel [110]$, measured with lower drive current on a different, much more resistive crystal in the same geometry.}
\end{figure}

The normal-state resistivity, shown in Fig.~\ref{fig:rho}, follows a low-temperature power law of $T^{3.67(6)}$ below $\sim$23\,K, and the residual resistivity ratio [RRR\,=\,$\rho$(300\,K)/$\rho$(0\,K)] of 375 suggests excellent crystal quality and low scattering.  Power laws above $T^2$ are observed in several metallic elements and alloys, most notably in silver\,\cite{Barber1975,Barnard1982} but also in some Al alloys\,\cite{Powell1960}. In particular, several closely-related tungsten and molybdenum aluminides have $T^4$ resistivity\,\cite{PeetsPRM2018}.  Such power laws have been attributed to an interplay of electron-electron, electron-phonon, electron-impurity, and electron-dislocation scattering\,\cite{Bergmann1980,Wiser1982,PeetsPRM2018}, but are seldom observed over such a wide range in temperature or to such high temperatures.  In the W and Mo aluminides, the relatively weak electron-electron interactions were considered as one possible reason for the lack of superconductivity, but \sixRe\ superconducts despite its similarly high power law.

A transition to zero resistivity, weakly suppressed by field, is tentatively observed at the same temperature as in the zero-field susceptibility and specific heat described later (lower inset in Fig.~\ref{fig:rho}), but the resistivity appears to partially recover at lower temperatures and low fields.  Due to noise issues, this transition was visible only in our most resistive samples, and it was present for a drive current of 10\,$\mu$A (current density 220\,A/m$^2$) but not 32\,$\mu$A (700\,A/m$^2$), suggestive of a very low critical current.  The extremely low resistivity in this material and this low current limit inevitably lead to poor data quality and prevent us from extracting firm conclusions on the superconductivity from resistivity data.  

\begin{figure}
  \includegraphics[width=\columnwidth]{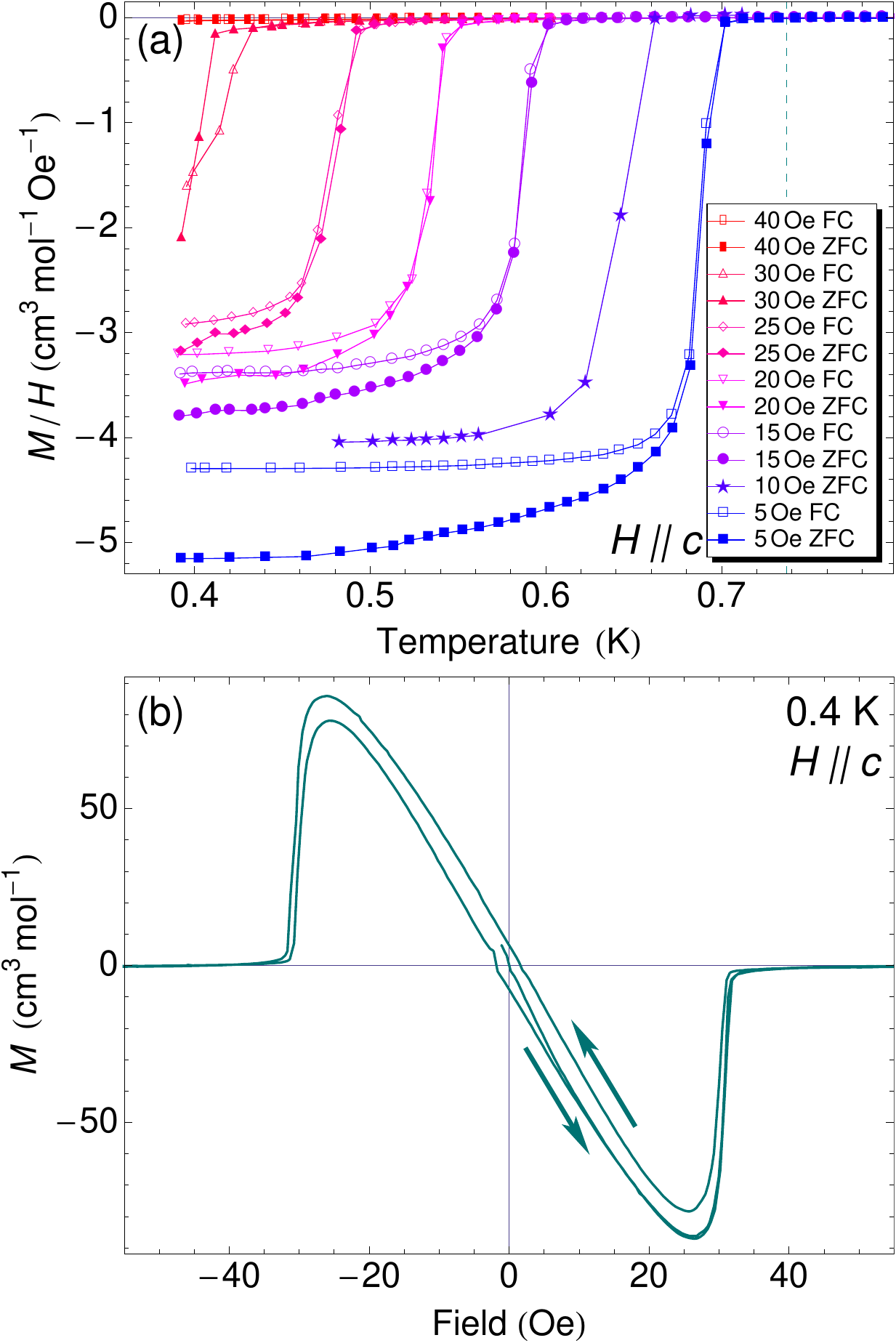}
  \caption{\label{fig:M}Magnetization data on \sixRe.  (a) Field-cooled and zero-field-cooled magnetization data in various fields applied along [001].  A dashed line marks the zero-field transition identified below by ac susceptometry. (b) $M$--$H$ loop, showing the rapid return to (nearly) zero magnetization associated with Type-I superconductivity.}
\end{figure}

\begin{figure*}[htb]
  \includegraphics[width=0.8\textwidth]{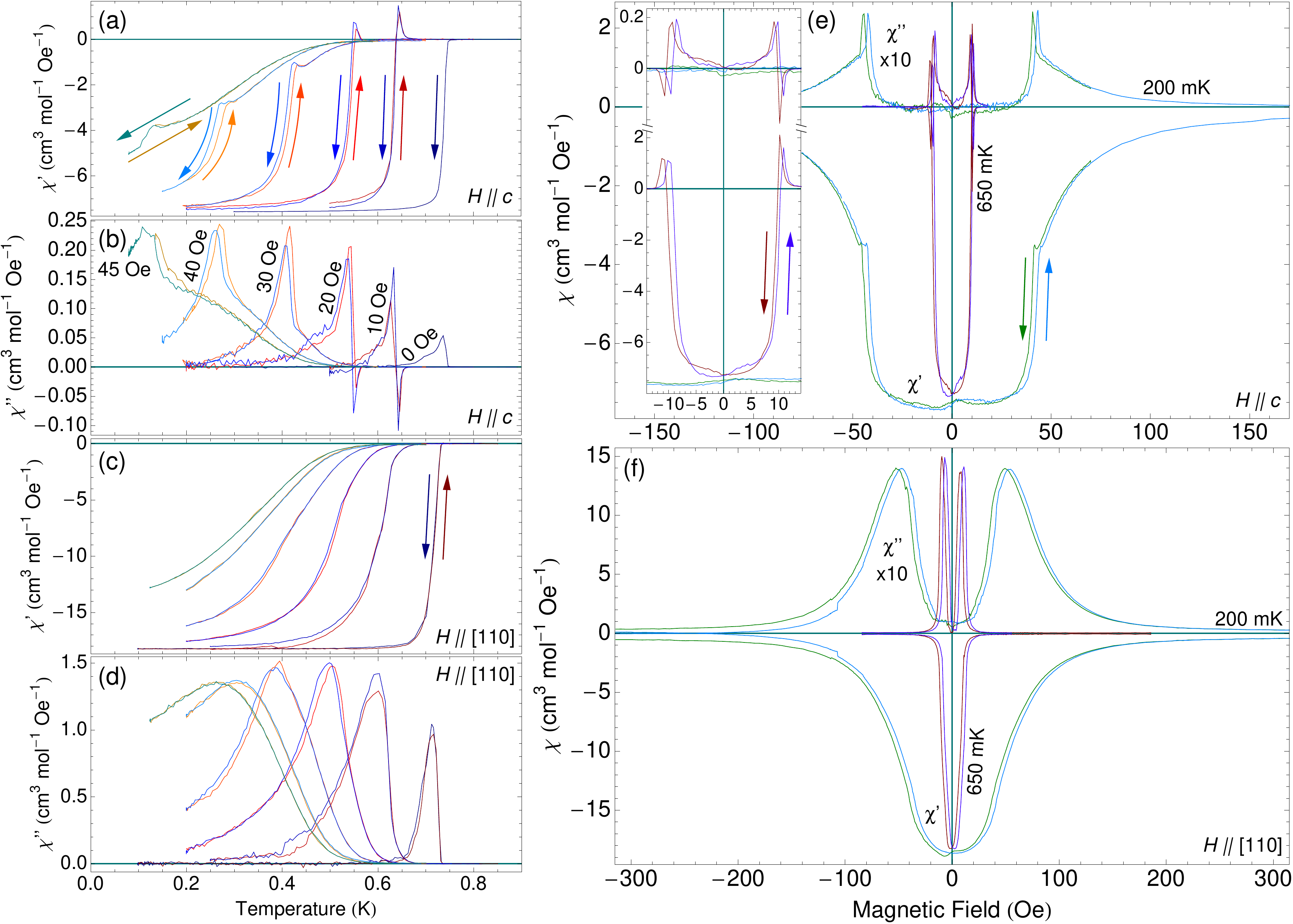}
  \caption{\label{fig:chi_all}ac Susceptometry of \sixRe.  (a) and (b) show the real and imaginary parts, respectively, of the temperature-dependent ac susceptibility in various fields $H||c$, while (c) and (d) show $H\perp c$.  Data were taken under zero-field-cooled warming (ZFC) and field-cooled cooling (FC) conditions.  (e) and (f) show field sweeps at 200 and 650\,mK for $H||c$ and $H\perp c$, respectively, with the imaginary components $\chi''$ expanded by a factor of 10.  The inset in (e) offers an expanded view at low field, with a vertical offset to better separate the real and imaginary components; here the scale for $\chi''$ has been expanded instead of the data.  The sample position in the measurement coil in this experiment is inexact and may vary between the two orientations --- $\chi$ values should not be considered quantitative.}
\end{figure*}

Magnetization data for various fields along the $c$ axis of a mosaic of several crystals are shown in Fig.~\ref{fig:M}(a).  These have not been corrected for demagnetization effects, a correction which is challenging given the unconventional shape of the crystals.  The zero-field-cooled magnetization in 5\,Oe reaches $\sim$95\%\ of full diamagnetic shielding, although note that this is inexact since demagnetization effects have not been considered, and the relatively small difference between field-cooled and zero-field-cooled data indicates a very small contribution from magnetic vortices.  This suggests Type-I superconductivity, which is confirmed by the $M$--$H$ loop shown in Fig.~\ref{fig:M}(b).  The magnetization is initially linear in field, then drops abruptly to nearly zero with no intervening vortex phase.  The fact that it does not drop to exactly zero may be due to thin-limit effects, and the broadening of the jump is a consequence of demagnetization effects.  

The low-temperature ac susceptibility of \sixRe\ is shown in Fig.~\ref{fig:chi_all}.  Sharp features indicating a transition are visible. This transition exhibits clear hysteresis in all field sweeps, and for temperature sweeps in finite field.  Additional hysteresis around zero field in the field sweeps is likely due to magnetic flux trapped either at the centre of the needle or in a surface layer.  It is noteworthy that in higher fields, the peak identifying the transition does not occur until after the material reaches a large fraction of its full zero-field suceptibility, and significant diamagnetic shielding survives to much higher fields or temperatures.  The unusual shape of the crystals suggests that demagnetization and thin limit effects may play a role here, but the similarity of the two field orientations is unexpected.

The shape of the transition is also unusual.  For nonzero fields applied along the needle, the real part of the susceptibility exhibits an initial positive spike.  This is not due to an extrinsic phase offset --- manually introducing such an offset leads to a physically unreasonable low-temperature $\chi''$ before having any significant effect on the spike itself.  This feature arises from the ``differential paramagnetic effect'', in which the rapid departure from (or return to) $M$=0 in a narrow window around the transition produces a $\partial M/\partial H$ that is large and positive\,\cite{Hein1961}.  It is also important that the magnetization process here be reversible and not subject to flux trapping or pinning.  This large positive slope at the transition is clearly visible in Fig.~\ref{fig:M}(b).  In principle, the differential paramagnetic effect is also possible in Type-II superconductors if $H_{c1}$ and \Hc2\ are very close, but the effect is strongly suggestive of Type-I superconductivity.

\begin{figure*}[htb]
  \includegraphics[width=\textwidth]{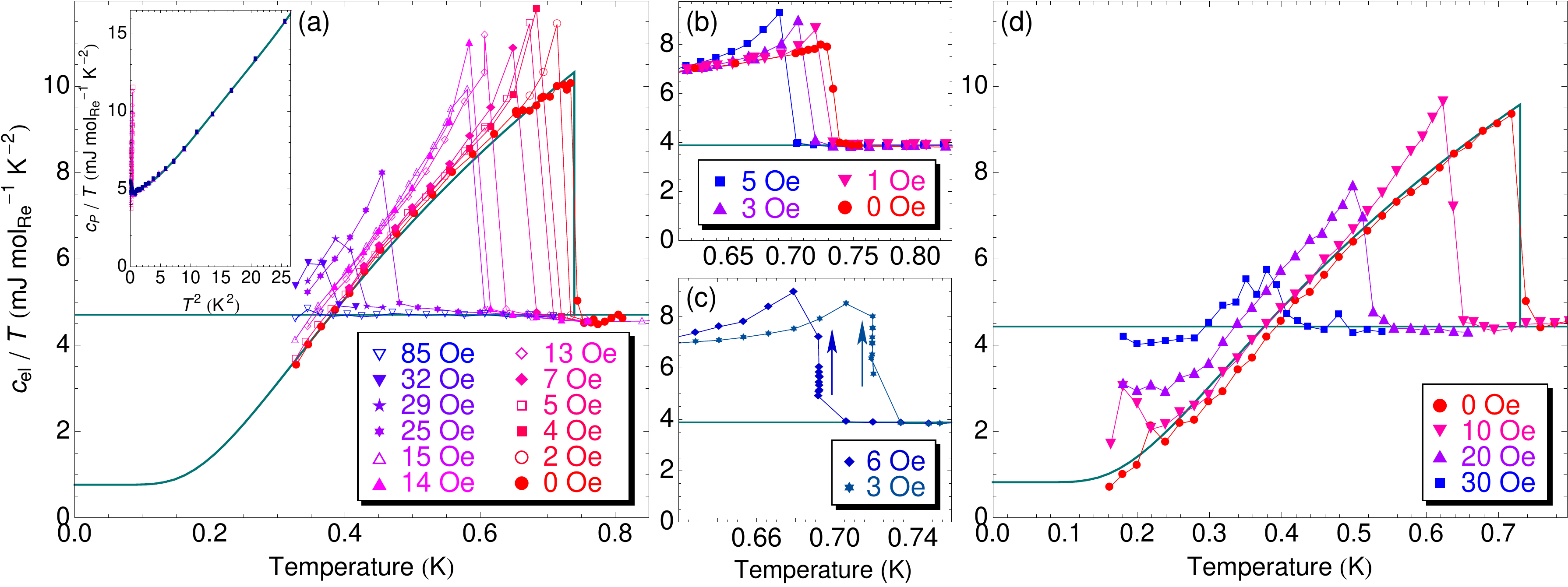}
  \caption{\label{fig:cP1}Electronic specific heat of \sixRe\ at various fields, divided by temperature to visualize the entropy balance.  The weak-coupling $s$-wave BCS expectation is superimposed, with an offset added to allow for a non-superconducting fraction.  The inset in (a) plots the specific heat at several fields as $c_P/T$ vs.\ $T^2$, showing the normal-state background that yields the electronic component.  (b) A first-order spike develops at the transition at very low fields, and (c) the data points collected for averaging at two fields increased monotonically or near-monotonically through the transition, presumably due to a slow relaxation process in the sample. (d) Data collected in a dilution refrigerator to resolve the low-temperature region remain consistent with BCS expectations, indicating a full gap.}
\end{figure*}

The electronic specific heat is compared against the weak-coupling $s$-wave BCS expectation in Fig.~\ref{fig:cP1}, with the total heat capacity and normal-state fit shown in its inset.  The small suppression relative to BCS [16\,\%\ in Fig.~\ref{fig:cP1}(a) and 19\,\%\ in Fig.~\ref{fig:cP1}(d)] and the nonlinearity in the background are attributed to a non-superconducting impurity phase on the surface, and were not observed in all mosaics measured.  The transition is extremely sharp, indicative of excellent sample quality, and we note that sharp transitions were consistently observed at essentially the same temperature despite the magnetic and specific heat measurements being performed on several distinct mosaics of 5--10 crystals. This indicates minimal variation between samples and suggests high consistency and homogeneity. Once the nonsuperconducting fraction is accounted for, the jump height at the transition $\Delta c_{el}/\gamma T_c$ is 1.37 in Fig.~\ref{fig:cP1}(a) and 1.42 in Fig.~\ref{fig:cP1}(d), consistent with the BCS weak-coupling expectation of 1.43.

The transition develops a spike in applied field, making the jump at the transition exceed the BCS expectation, rather than shrinking as is normally observed in Type-II superconductors.  Higher-resolution data, shown in Fig.~\ref{fig:cP1}(b), confirm this result and demonstrate that the spike appears even for fields as low as $\sim$2\,\%\ of $H_c$.  Such a spike indicates a latent heat, implying a first-order transition in field, which is characteristic of Type-I superconductors.  Below the transition, the specific heat of a Type-I superconductor is essentially field-independent, because the vast majority of the sample feels zero field and the Meissner supercurrents at the edge carry zero entropy.  The in-field specific heat rapidly returns to its zero-field value after a narrow spike at the transition.  As field is increased, this large spike moves down in temperature, initially growing in area, then ultimately shrinking, to ensure entropy balance between the normal and superconducting states at the transition.  This was observed in two fields in thallium\,\cite{Keesom1934,Keesom1934b}, and later in Al\,\cite{Rorer1963}, and is also seen in most modern studies on Type-I compound superconductors, listed below.  As part of this study, we also measured the specific heat of $\beta$-Sn for comparison; these data are shown in Appendix~\ref{Sn}.  In contrast, in a Type-II superconductor above its lower critical field, the introduction of normal-state vortex cores within the superconducting bulk leads to non-zero specific heat at low temperatures and a reduced jump at the transition.  As field is applied, this progressively reduces the deviations from the normal state specific heat.  The specific heat of \sixRe\ in low fields is characteristic of Type-I superconductivity.

The low-temperature specific heat of \sixRe\ in field was remeasured in a dilution refrigerator to distinguish these possibilities, as shown in Fig.~\ref{fig:cP1}(d) --- the low-temperature data do eventually depart from the low-field BCS form, but this is likely attributable to broadening of the transition as a consequence of the unusual sample shape (this also happens in $\beta$-Sn, as seen in Fig.~\ref{fig:Sn}.  The low-temperature behavior in low field indicates a full superconducting gap.  

At two fields, a temperature point happened to coincide with the transition [Fig.~\ref{fig:cP1}(c)].  These temperature scans were performed on cooling, ten measurements were averaged at each temperature near the transition, and the specific heat was found to increase with time when the temperature coincided with the transition.  The relaxation time method oscillates the temperature slightly, encouraging relaxation of the sample, the coupling between the sample and the measurement platform was consistently excellent, and the measurements at the transition took a total of $\sim500\times$ the relaxation time constant to complete --- this slow relaxation is additional strong evidence for a latent heat.  

From the specific heat datasets, a number of parameters characterizing the superconducting and normal states can be extracted.  The Sommerfeld coefficient $\gamma$ describing the electronic specific heat is 3.9-4.7\,mJ/mol$\cdot$K, and the low-temperature Debye temperature describing the phonon contribution, $\Theta_D$, is $\sim$400\,K.  Extrapolating $c_P/T$ to $T=0$ and integrating yields a thermodynamic critical field of 43.3\,Oe, not far from the observed critical field of $\sim$44-50\,Oe.  It would be possible to estimate many additional parameters if we could assume a carrier concentration and a single spherical Fermi surface.  However, since at least four bands cross the Fermi level according to the online band structure calculation repositories (based on density functional theory) at the Materials Project\,\cite{MaterialsProjectAl6Re,Jain2013} and Materiae\,\cite{MateriaeAl6Re}, such estimates would likely be highly inaccurate.

\section{Discussion}

Since signatures of Type-I behaviour are observed in the magnetization, ac susceptibility, and specific heat; and since the thermodynamic $H_c$ calculated from the zero-field specific heat is in reasonable agreement with that extracted from measurements in field, we conclude that \sixRe\ is a Type-I superconductor. Most of the pure elemental superconductors are Type-I, but so few compounds are known to be Type-I that they can be readily enumerated: AuAl$_2$ and AuIn$_2$\,\cite{Soulen1977,VanVechten1983}, K-intercalated graphite in certain regimes\,\cite{Kobayashi1981}, \{Y,La,Lu\}Pd$_2$Si$_2$\,\cite{Palstra1986}, LaRh$_2$Si$_2$\,\cite{Palstra1986}, YbSb$_2$\,\cite{Yamaguchi1987,Zhao2012}, TaSi$_2$\,\cite{Gottlieb1992}, Ag$_5$Pb$_2$O$_6$\,\cite{Yonezawa2005}, B-doped SiC\,\cite{Ren2007,Muranaka2008,Kriener2008,Kriener2008b}, \{Sc,Lu\}Ga$_3$\,\cite{Svanidze2012}, SnAs\,\cite{WangSnAs2014}, KBi\,\cite{Sun2016}, AuBe\,\cite{Rebar2019,Singh2019,Beare2019}, and probably LaPd$_2$Ge$_2$\,\cite{Hull1981} and LaRhSi$_3$\,\cite{Anand2011}. We observe that many of the known Type-I superconductors were only discovered within the last decade or so.  This may suggest that their apparently scarcity is a consequence of many intermetallic superconductors being discovered and characterized before the distinction between Type-I and Type-II was known, while a renewed focus recently on intermetallics as potential noncentrosymmetric superconductors or as hosts of topological electronic matter has recently uncovered most of the known examples.  

\begin{figure}[htb]
  \includegraphics[width=\columnwidth]{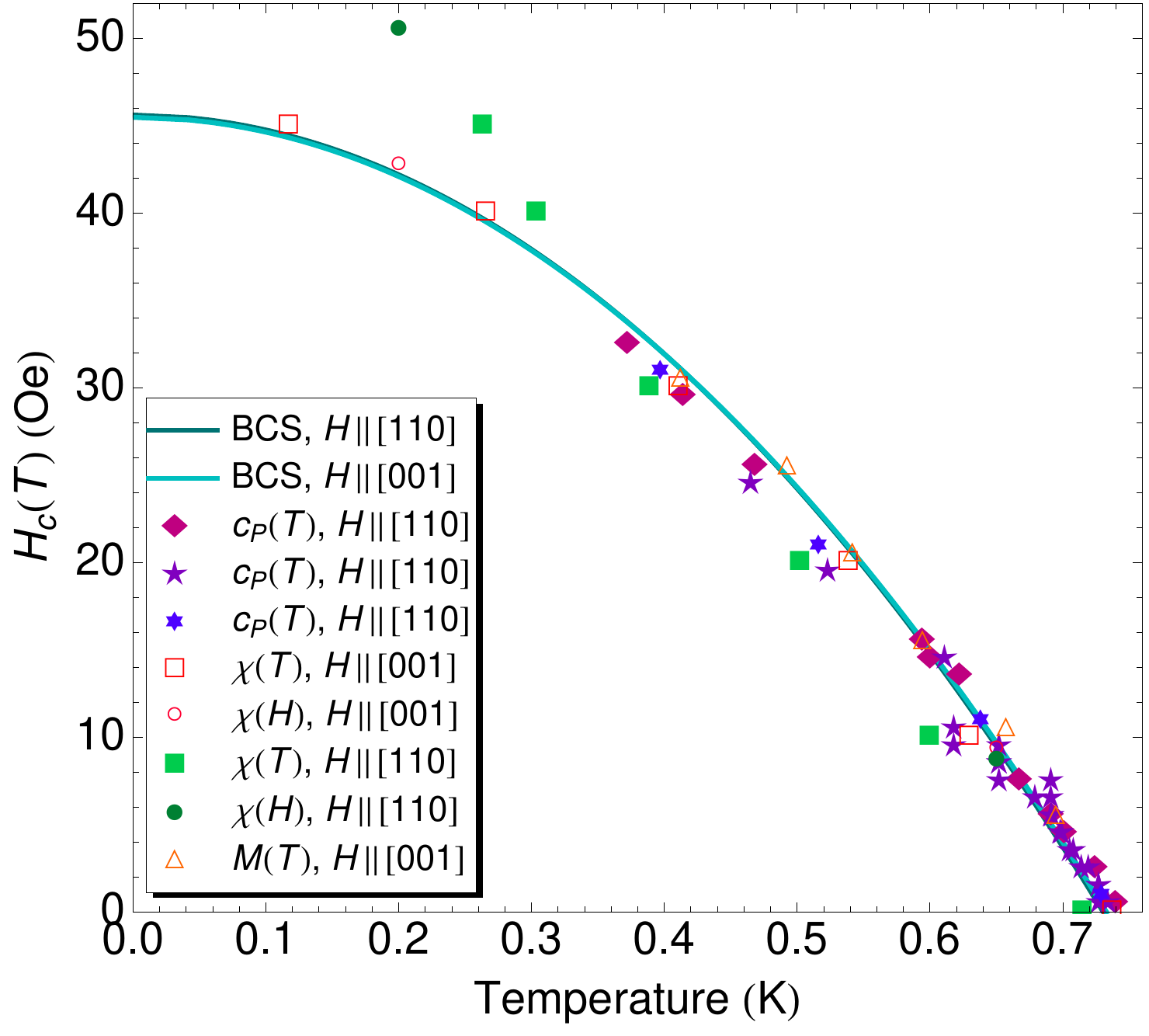}
  \caption{\label{fig:HT}$H$--$T$ phase diagram extracted from specific-heat, magnetization, and ac-susceptibility measurements, based on an entropy-conserving construction, 10\% of full diamagnetism, and an average of the peaks in $\chi''$ for all sweep directions, respectively.}
\end{figure}

The $H$--$T$ phase diagram extracted from the transitions in the magnetization, susceptometry and specific heat is shown in Fig.~\ref{fig:HT}.  Aside from the lowest-temperature points for fields along [110], the shape is consistent with the BCS form, $H_c(T)=H_c(0)\sqrt{1-(T/T_c)^2}$.  These low-temperature points are extracted based on the peak position of a very broad and asymmetric transition, and have the largest uncertainties of any points in the phase diagram.  Despite the inclusion of these three points, the BCS fits are nearly identical for the two field orientations, resulting in a critical field of 48\,Oe.  We are not able to resolve any anisotropy.  Resistivity data are not included here due to difficulties in performing the measurement, identifying a field offset, and reliably determining the point at which zero resistivity is attained.  However, we note that the field dependence in the resistivity would appear to be weaker by an order of magnitude or more.  This could arise from portions of the sample being in the thin limit, either due to the sample shape or due to superconductivity surviving at surfaces or grain boundaries.  Superconductivity surviving in the thin limit at twin boundaries has been suggested as a possible explanation for similar order-of-magnitude differences between the resistive and bulk upper critical fields in BiPd\,\cite{Peets2016}.

One remaining question is why \sixRe\ superconducts while the W and Mo aluminides studied previously\,\cite{PeetsPRM2018} do not.  All materials have a low-temperature resistivity power law around $T^4$, so scattering processes and electron-phonon coupling strength should be similar.  Unlike the W and Mo materials, however, \sixRe\ is centrosymmetric.  We speculate that this makes it easier for Cooper pairs to form. In the absence of magnetism, the pairing would presumably be phonon-based, in which pairing with predominantly singlet character (i.e. $1/\sqrt2\left[\left|k_\uparrow\right\rangle\left|-k_\downarrow\right\rangle - \left|k_\downarrow\right\rangle\left|-k_\uparrow\right\rangle\right]$) and with minimal momentum-dependence generally offers the best energy savings upon entering the superconducting state. Inversion symmetry enforces parity, so performing inversion about the origin in momentum space can at most change an overall sign, even in the presence of spin-orbit coupling.  Without inversion, however, the bands will in general be spin-split as a result of spin-orbit coupling.  This may mean, for instance, that the $\left|k_\uparrow\right\rangle\left|-k_\downarrow\right\rangle$ and $\left|k_\downarrow'\right\rangle\left|-k_\uparrow'\right\rangle$ components are at different momenta $k$ and $k'$ (or one may be entirely absent), making a singlet combination higher-energy or impossible to construct. Re also differs in electron count.  As mentioned previously, one key Matthias Rule states that having six valence electrons is bad for superconductivity, while having five or seven provides the highest transition temperatures\,\cite{Matthias1955}.  Metallic W and Mo have six, while Re has seven, although it is questionable whether such a rule should apply in a material that is almost entirely Al. Determining the relative importance of these two considerations would require studying additional members of this family, either centrosymmetric W or Mo aluminides or noncentrosymmetric Re aluminides.

\section{Conclusion}

Unlike a family of related W and Mo aluminides, \sixRe\ superconducts.  Even in low field, it exhibits hysteresis in its magnetic properties, a spike at the transition in the specific heat indicating a latent heat, and extremely long relaxation processes at the transition.  This combination constitutes compelling evidence for a first-order superconducting phase transition in field, and taken together with the step-like $M$--$H$ curve and resulting differential paramagnetic effect, offers strong evidence for Type-I superconductivity, making \sixRe\ the newest of a very small group of compound Type-I superconductors.  The chief differences most likely responsible for \sixRe\ being a superconductor while the W and Mo materials are nonsuperconducting are that the latter are noncentrosymmetric, and Re has a valence electron count that is more favorable for superconductivity.  Further investigation of this family would be required to establish to what degree each of these effects is responsible for \sixRe\ being a superconductor.  The large number of such aluminides, readily grown from Al flux with transition metal ions from columns VIB and VIIB, makes this family particularly well suited to a detailed investigation of the interplay of these effects.  This would be particularly useful for determining the extent to which a lack of spatial inversion suppresses superconductivity.

\begin{acknowledgments}
The authors thank Yue-Jian Lin of the Shanghai Key Laboratory of
Molecular Catalysis and Innovative Materials for assistance with
single crystal x-ray diffraction, structure solution, and
refinement. This work was supported by the National Key R\&D Program
of the MOST of China (Grant No.~2016YFA0300203), the Science Challenge
Program of China (Grant No.~TZ2016004), the National Natural Science
Foundation of China (Projects No.~11650110428, 11421404, U1532266,
11790312, and 11674367), and the Zhejiang Provincial Natural Science
Foundation (Grant No.~LZ18A040002). DCP is supported by the Chinese
Academy of Sciences through 2018PM0036, and MK is supported by a
Grant-in-Aid for Scientific Research from the Japan Society for
the Promotion of Science (No.~15K05140).
\end{acknowledgments}

\bibliography{Al4W}

\appendix\clearpage
\section{Crystal Structure Refinement Details\label{XRDeets}}

Details of the crystal structure refinement of \sixRe\ are described in Tab.~\ref{tab:XRD}, and its anisotropic displacement parameters are provided in Tab.~\ref{tab:U}.  Further details can be extracted from the crystallographic information file (CIF), available online as ancillary information\,\cite{PeetsArXiv2019Supp}.

\begin{table}[htb]
  \caption{\label{tab:XRD}Details of the structure refinement of \sixRe\ at 203(2)\,K.}
  \begin{tabular}{ll}\\ \hline\hline
    Formula & Al$_6$Re\\
    Space group & $Cmcm$ (\#~63)\\
    $a$ & 7.599(4)\,\AA\\
    $b$ & 6.602(4)\,\AA\\
    $c$ & 9.040(5)\,\AA\\
    $Z$ & 4\\
    $F(000)$ & 612\\
    $\theta$ range & 4.089 to 27.999$^\circ$\\
    Index ranges & $-$10$\leq$$h$$\leq$8, $-$7$\leq$$k$$\leq$8, $-$11$\leq$$l$$\leq$10\\
    Total reflections & 1634\\
    Independent reflections & 314\\
    Goodness of fit & 1.096\\
    $R$ factors, all data & $R_1$=0.0200, $wR_2$=0.0376\\
    $R$ factors, $I$$>$2$\sigma(I)$ & $R_1$=0.0165, $wR_2$=0.0371\\
    Extinction coefficient & 0.0154(7)\\ \hline\hline
  \end{tabular}
\end{table}
    
\begin{table}[htb]
  \caption{\label{tab:U}Anisotropic displacement parameters (\AA$^2$) for \sixRe\ in $Cmcm$.  The anisotropic displacement factor exponent takes the form $-2\pi^2\left[h^2a^{*2}U_{11}+ \text{...} +2klb^*c^*U_{23}\right]$.}
  \begin{tabular}{lr@{.}lr@{.}lr@{.}lr@{.}lcr@{.}l}\hline\hline
    Site & \multicolumn{2}{c}{$U_{11}$} & \multicolumn{2}{c}{$U_{22}$} & \multicolumn{2}{c}{$U_{33}$} & \multicolumn{2}{c}{$U_{12}$} & $U_{13}$ & \multicolumn{2}{c}{$U_{23}$} \\ \hline
    Re & 0&007(1) & 0&004 & 0&005(1) & 0& & 0. & 0& \\
    Al(1) & 0&010(1) & 0&012(1) & 0&009(1) & 0& & 0. & 0&000(1) \\
    Al(2) & 0&011(1) & 0&009(1) & 0&012(1) & 0& & 0. & 0&003(1) \\
    Al(3) & 0&010(1) & 0&006(1) & 0&014(1) & 0&002(1) & 0. & 0& \\ \hline\hline
  \end{tabular}
\end{table}

\section{Specific heat of Sn\label{Sn}}

\begin{figure}[htb]
  \includegraphics[width=\columnwidth]{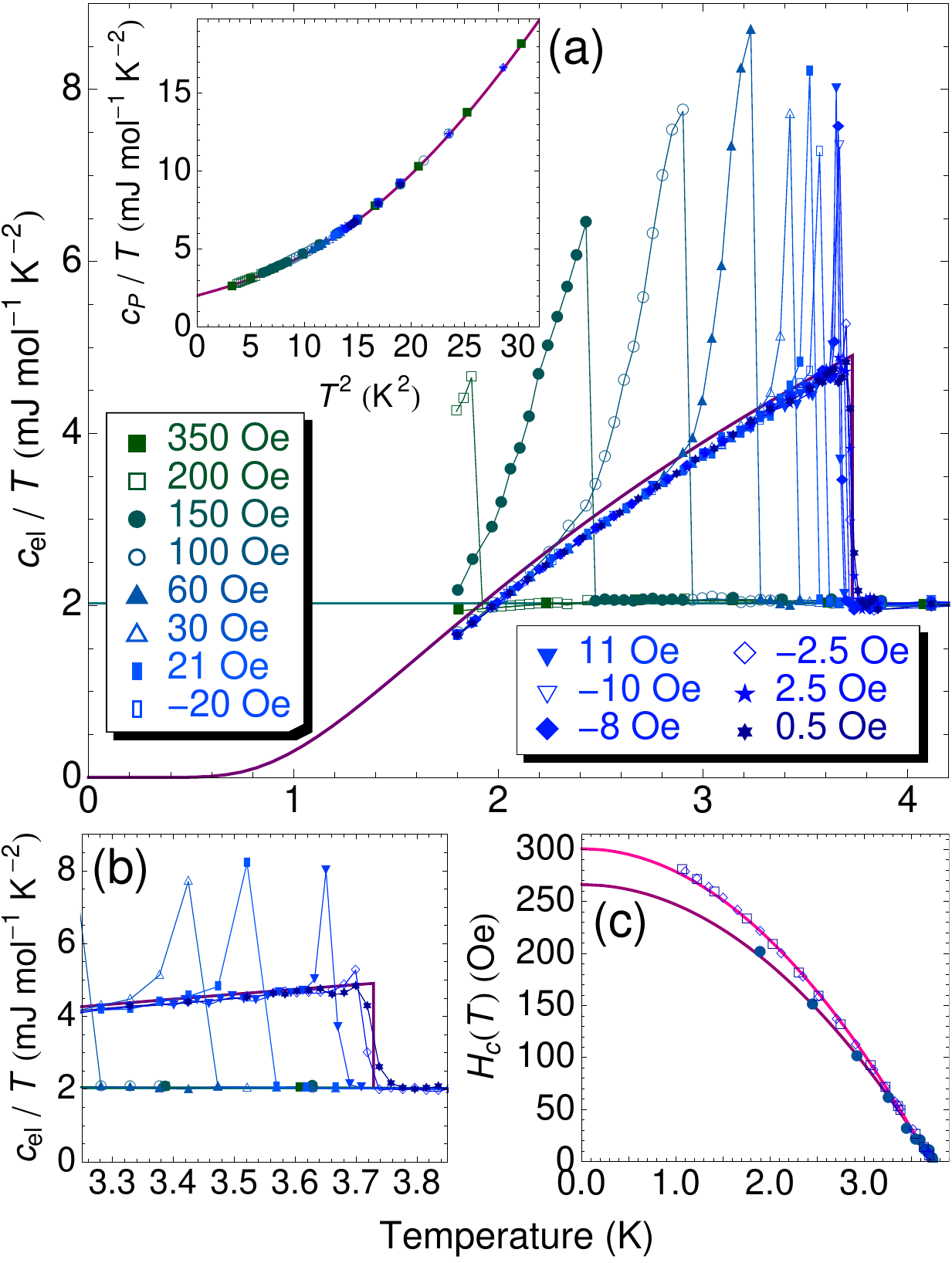}
  \caption{\label{fig:Sn}Specific heat of $\beta$-Sn in field.  (a) Electronic specific heat of $\beta$-Sn in various fields. The inset shows the normal-state specific heat, used as a baseline to extract the electronic specific heat, based on all data taken above $H_c(T)$.  (b) Expanded view of selected data sets near the zero-field \Tc\ [same symbols and colors as in (a)]. (c) $H$--$T$ phase diagram extracted for $\beta$-Sn from the specific heat transitions. Hollow data points are taken from ballistic induction measurements reported for two samples in Ref.~\onlinecite{Shaw1960}.}
\end{figure}


The specific heat of $\beta$-tin was measured as a Type-I superconducting standard for comparison.  A hemispherical sample of $\beta$-Sn was obtained by cutting a piece of tin shot from Alfa Aesar (99.999\% pure, polycrystalline) in half.  Specific heat was measured in a PPMS using the relaxation-time method, with the sample mounted to a $^4$He specific heat puck using Apiezon N grease.  The sample was maintained well above the $\alpha$-$\beta$ phase transition until cooling for the measurement, and no sign of the gray $\alpha$ phase was visible either before or after the measurement.  The specific heat is shown in Fig.~\ref{fig:Sn}(a), an expanded view of selected fields near the zero-field transition is shown in Fig.~\ref{fig:Sn}(b), and the resulting $H$--$T$ phase diagram is shown in Fig.~\ref{fig:Sn}(c).  
  
\end{document}